\documentclass[10pt,a4paper]{article}
\usepackage[utf8]{inputenc}
\usepackage{amsmath}
\usepackage{amsfonts}
\usepackage{amssymb}
\usepackage{hyperref}
\newcommand{\nn}{\nonumber}
\begin{document}
\title{NC plane waves, Casimir effect and flux tube potential with L\"uscher terms}
\author{Samuel Kov\'a\v{c}ik, Peter Pre\v{s}najder}
\maketitle
\abstract{We analyze plane waves in a model of quantum mechanics in a three dimensional noncommutative (NC) space $\textbf{R}^3_\lambda$. Signature features of NC models are impossibility of probing distances smaller than a certain length scale $\lambda$ and a presence of natural energetic cut-off at energy scale of order $\lambda^{-2}$ (in convenient units). We analyze consequences of such restrictions on a 1 dimensional Casimir effect. The result shows resemblance to flux tube potential for $q\bar{q}$ pairs and to effective bosonic string theories with L\"usher terms. Such behavior might effect the radius of possible compact (fuzzy) dimensions.}
\section{Introduction}
NC models are believed to play an important role in a theory merging quantum theory with gravity. If we tried to probe distances of the order of Planck length the probing particle (for example a photon) would be hidden under its event horizon, a black hole would be formed and no information would be obtained. Because of this we expect theory of quantum gravity to have a nontrivial commutator of the position operators $[x,x]\neq 0$ which would forbid an exact position measurements. By removing arbitrarily small distances we also remove arbitrarily large energies and the theory obtains a natural energy cut-off. 

This should be by no means a complete construction of quantum mechanics in noncommutative space (NC QM), quite the opposite, we will try to be as brief as possible. For a detailed construction we refer the reader to \cite{us1, us2, us3, us4} where the current model was developed and used to analyze many different problems. 

The noncommutativity of space coordinates is described by a relation
\begin{equation}\label{com}
 [x_i,x_j]\ =\ 2i\,\lambda\,\varepsilon_{ijk}\,x_k\,,
\end{equation}
Such NC coordinates can be realized with two sets of auxiliary bosonic operators (lacking any physical interpretation so far) satisfying
\begin{equation}\label{aux}
[a_\alpha,a^\dagger_\beta]\,=\,\delta_{\alpha\beta },\ \
[a_\alpha,a_\beta]\,=\,[a^\dagger_\alpha, a^\dagger_\beta]\,=\,0\, ; \, \alpha , \beta = 1,2 \, ,
\end{equation}
that act in an auxiliary Fock space $\mathcal{F}$ spanned on normalized vectors
\begin{equation}
|n_1,n_2\rangle\ =\ \frac{(a^\dagger_1)^{n_1}\,(a^\dagger_2)^{n_2}}{
\sqrt{n_1!\,n_2!}}\ |0\rangle\,.
\end{equation}
The NC coordinates are defined using Pauli matrices as
\begin{equation}\label{x}
 x_j\ =\ \lambda\,a^+\,\sigma_j\,a\ \equiv\
\lambda\,\sigma^j_{\alpha\beta}\,a^\dagger_\alpha\,a_\beta,\
j\,=\,1,2,3 \, .
\end{equation}

The Hilbert space $\mathcal{H}$ consist of states of form
\begin{equation} \label{Psi}
\Psi(\vec{x}) = \Psi (a_1, a_2,a_1^+,a_2^+) = \sum\, C_{m_1 m_2 n_1 n_2}\,(a^\dagger_1
)^{m_1}\,(a^\dagger_2)^{m_2}\,(a_1)^{n_1}\,(a_2)^{n_2} \, ,
\end{equation}
and is equipped with a (trace) norm 
\begin{equation} \label{tn}
\| \Psi \|^2\ =\ 4\pi\,\lambda^3\,\mbox{Tr} [\Psi^\dagger\,(N+1)\,\Psi]\ =\ 4\pi\, \lambda^2\,\mbox{Tr}[\Psi^\dagger\,r\,\Psi] \, ,
\end{equation}
which in the commutative limit ($\lambda \rightarrow 0$) reproduces the ordinary integral norm (the factor $4 \pi \lambda^2 r$ is chosen to ensure that). 

For what follows it is sufficient to know following operators on $\mathcal{H}$: the coordinate operator, the free Hamiltonian and the velocity operator
\begin{eqnarray}
\hat{X}_i \Psi &=& \frac{1}{2}\left( x_i \Psi + \Psi x_i \right) \, , \nonumber \\
\hat{H}_0 \Psi &=& \frac{1}{2\lambda r} [ a^+_\alpha, [ a_\alpha , \Psi ]] \, , \nonumber\\
\hat{V}_i \Psi &=& -i [ \hat{X}_j , \hat{H}_0]\Psi =\frac{i}{2r} \sigma^i_{\alpha \beta} (a^+_\alpha \Psi a_\beta - a_\beta \Psi a^+_\alpha) \, .
\end{eqnarray}

As has been showed in \cite{us2}, these operators mirror the behavior of corresponding operators in ordinary QM up to a possible $\lambda$ dependent corrections. We call $\hat{V}_i$ a velocity operator, since our space lacks the translational invariance, but in variety of cases it is equal to the momentum operator (with $m=1$) and the gradient operator (up to normalization). In the same reference it has been showed that 
\begin{equation} \label{VVH}
\left( \frac{1}{\lambda}\,-\,\lambda\,\hat{H}_0 \right)^2 \ =\
%\frac{1}{\lambda ^2} \left( \right)
\frac{1}{\lambda^2} \,-\,\hat{V}^2_j  \,.
\end{equation}
which demonstrates the free energy cut-off $E \in \langle 0,2\lambda^{-2}\rangle$. We define 
\begin{equation}
\hat{V}_4\,\Psi\ =\ \left( \frac{1}{\lambda}\,-\,\lambda\, \hat{H}_0\right)\,\Psi\ =\ \frac{1}{2r} \left( a^+_\alpha\, \Psi\, a_\alpha\, +\, a_\alpha\, \Psi\, a^+_\alpha \right)
\end{equation}
for which it holds that
\begin{equation} \label{Va2}
\hat{V}^2 + \hat{V}_4^2 = \lambda^{-2} \, .
\end{equation}

This is perhaps all we have to be (light)equipped with to search for NC plane waves.

\section{NC plane waves and the Casimir effect}

\textbf{NC plane waves:} In the ordinary QM mechanics planewaves satisfy $\partial_i \Psi_\textbf{k} \propto k_i \Psi_\textbf{k}$ where $k_i$ are constant. This can be easily transferred to NC QM as
\begin{equation} \label{eigenE}
\hat{V}_i \Psi_\nu = \nu_i \Psi_\nu \, ,
\end{equation}
without a loss of generality we can set $\nu_1=\nu_2=0, \, \nu_3 = \nu$. The remaining equation $\hat{V}_3 \Psi_\nu = \nu_3 \Psi_\nu$ can be solved 'from the scratch' by methods developed in \cite{us4}. However, solving equations can be simplified by an educated guess and it is not very hard to make one here, we will seek the solution of the form $e^{i q x_3}$ (note there is a different constant in the exponent as in \eqref{eigenE}). Since $x_3 = \lambda \left( a^+_1 a_1 - a^+_2 a_2\right)$ we can get use of the fact following from \eqref{aux} that (using notation $N_1 = a^+_1 a_1, \ N_2 = a^+_2 a_2$) $f(N_1) a_1 = a_1 f(N_1-1), \, f(N_1)a^+_1 = a^+_1 f(N_1+1)$ (same for the second set) to show that
\begin{eqnarray} \label{eigenV}
\hat{V}_3 \Psi_\nu &=& \frac{i}{2r} \left( a^+_1 e^{i \lambda q (N_1 -N_2)}a_1 - a^+_2 e^{i \lambda q (N_1 -N_2)} a_2 \right) \\ \nn
&&- \frac{i}{2r}\left( a_1 e^{i \lambda q (N_1 -N_2)} a^+_1 - a_2 e^{i \lambda q (N_1 -N_2)}a^+_2 \right) \\ \nn
&=&\frac{i}{2r}\left(a^+_1a_1 e^{i \lambda q (N_1-1 -N_2)} +a^+_2 a_ e^{i \lambda q (N_1 -N_2+1)} \right) \\ \nn
&&-\frac{i}{2r}\left( a_1 a^+_1 e^{i \lambda q (N_1+1 -N_2)}-a_2 a^+_2 e^{i \lambda q (N_1 -N_2-1)} \right) \\ \nn
&=& \frac{i}{2\lambda}\left( e^{i \lambda q (N_1 -N_2 -1)-e^{i \lambda q (N_1 -N_2+1)}}\right) = \frac{\sin \lambda q}{\lambda} \Psi_\nu
\end{eqnarray}

The eigenvalue is $\lambda^{-1} \sin \left( \lambda q \right)$ instead of just $q$ (they are equal in the $\lambda \rightarrow 0$ limit). By the same steps it can be shown that 
\begin{equation} \label{ene}
 \hat{V}_4 \Psi_\nu = \left( \frac{1}{\lambda}\,-\,\lambda\, \hat{H}_0\right) \Psi _\nu = \frac{\cos \lambda q}{\lambda} \Psi_\nu \, ,
\end{equation}
and \eqref{Va2} follows directly. 

In the region where $q \lambda \ll 1$ the spectrum is nearly identical to the ordinary case, it differs considerably for $q \lambda \approx 1$. The maximal value is obtained for $q = \frac{\pi}{2\lambda}$. \\
\textbf{1 dimensional Casimir effect:} The minimum energy of a quantum oscillator is $\frac{1}{2}\hbar \omega$ and the ground state of a system has the total energy equal to sum of all zero mode frequencies. The result depends on the geometry of the system (both shape and size) and therefore a force is exerted on it. Such force is called Casimir and has been measured for example between parallel metallic surfaces \cite{casExp}. A thorough review of this effect can be found for example in \cite{cas}.
The most simple problem to investigate consequences of \eqref{eigenV} is perhaps a Casimir effect in 1 dimension. In the commutative case, using zeta-regularization (and $\zeta(-1) = - \frac{1}{12}$), one obtains after summing over all zero modes (and 2 polarizations) of EM fields on a line segment (of length $L$) with possible frequencies  $q_n = \frac{n \pi}{L}$  the total energy (using convenient units and taking into account outside fields as well)
\begin{equation} \label{res0}
E_{0}(L)=\sum \limits_{n=1}^\infty \frac{n \pi}{L} = - \frac{\pi}{12 L} \, .
\end{equation}

We shall now study how does replacing the ordinary frequencies with NC ones $q \rightarrow \lambda^{-1} \sin \left( \lambda q \right)$ amends this result. After summing over all possible frequencies \footnote{Assuming $L/2\pi$ is an integer, which is a good approximation since $\lambda$ is expected to be a fundamental length.} (without taking outside fields into account)
\begin{equation} \label{res1}
E_{0,\lambda}(L) = \sum \limits_{n=1}^{\frac{L}{2\lambda}} \frac{\sin \frac{n \pi}{L} \lambda}{\lambda} = \frac{1+\cot \frac{\pi \lambda}{2 L}}{2 \lambda} \, ,
\end{equation}
where in the upper summation range the energy (or frequency) achieves its maximum. Taylor expansion of the result around $L=0$ is
\begin{equation} \label{res2}
E_{0,\lambda}(L) = \frac{L}{\pi \lambda^2}+\frac{1}{2\lambda} - \frac{\pi}{ 12 L} - \frac{\pi^3 \lambda^2}{720 L^3}+O(\lambda^4) \, .
\end{equation}

Let us make several comments here. To begin with an obvious one – the result is finite and in the $\lambda \rightarrow 0$ limit reproduces \eqref{res0} (the singular part gets canceled by outside fields). \eqref{res2} provides a good interpretation of the enigmatic result $1+2+3+... = -\frac{1}{12}$. It should be looked at as $1+2+3+... = \infty - \frac{1}{12}$ (as can be seen by taking $\lambda \rightarrow 0$ in \eqref{res2}).

There are two possible scenarios to be distinguished: either the fields exist only on the line (for example if the space is compact and the line is all there is, or the fields are suppressed elsewhere), or also outside (for example 1 dimensional analogue of the case of parallel conductive plates). If energy of the outside fields is taken into account, then the linear term gets canceled. In such case the system consists of three line segments $\langle -\Lambda, a \rangle, \ \langle a, b \rangle, \ \langle b , \Lambda \rangle$, where in the end we take $\Lambda \rightarrow \infty$. The contribution from linear terms is $\left( \pi \lambda^2 \right)^{-1} \left( \left(a - \left( - \Lambda\right) \right) + \left( b-a \right) + \left( \Lambda - b\right) \right) = \left( \pi \lambda^2 \right)^{-1} 2 \Lambda$ which is a infinity independent of $a,b$ in the considered limit. Only other contribution from the outside fields is again an infinite constant $\frac{1}{2\lambda}$, since all other terms vanish as some power of $\Lambda^{-1}$.

If the linear term persists it becomes dominant in $L \gg \lambda$ region and leads to a confinement(-like behavior). This term is proportional $\lambda^{-2}$, which corresponds to the scale of the energetic cut-off (perhaps the Planck energy). The constant term generates the $L$ independent infinity which was taken care of by the zeta-regularization in \eqref{res0}. All other terms are of the form $ - \frac{\lambda^{2i}}{L^{2i+1}}$ with $i = 0,1,2,...$.

The energy reaches its minumum for $\frac{\pi \lambda}{2 L} = \pi$, however this value probably cannot be achieved, in this region is the sum \eqref{res1} ill-definied (the upper boundary is $1/4$. The largest achievable frequency is equal to $\lambda^{-1}$ and therefore the shortest possible wavelength is $\Lambda_{\mbox{min}}= 2 \pi \lambda$. Below $L= \frac{1}{2}\Lambda_{\mbox{min}} = \pi \lambda$ there cannot be squeezed even the shortest possible wavelength. The attractive force vanishes at the length scale $L\approx \lambda$.

\section{Conclusion}

We have derived the form of NC QM plane waves, whose spectrum strongly differs from that of ordinary theory, what corresponds to natural energetic cut-off and a maximal obtainable frequency. This affects many physical problems, we have chosen the Casimir effect to demonstrate it. 

The linear terms (if it prevails) leads to a confinement like behavior. In the case of compact extra dimensions this should lead to shrinking of the radius to $R \approx \lambda$, where significance of other (Casimir) terms grows as well. Similar behavior appears in $q\bar{q}$ pairs where color screening creates a flux tube with a linear potential on a line. In this case $\lambda^{-1}$ should be connected to a mass scale of the theory.

There has been analyzed a possible connection between $q\bar{q}$ flux tubes and (effective) string theories. Simulations and calculations suggest that the linear (confinement) term should be supplemented by an additional terms, first of which is proportional to $L^{-1}$ (where $L$ is now the string length). Let us quote the results of L\"usher et al. in  \cite{lusher1}
\begin{equation}
E(L) = T L + C - \frac{\gamma}{L} + ... 
\end{equation}
$T$ is the string length (which is related to the Planck's scale), $C$ is an arbitrary mass factor (it is sometimes referred to as the regulazation term) and $\gamma= \frac{\pi}{12}$. This seems to be in an exact agreement with \eqref{res2}. In \cite{lusher2} it has been argued that no $L^{-2}$ term should appear, which again agrees with \eqref{res2}, but that next term should be $-\frac{\pi^3 \lambda^2}{288 L^3}$, where the numerical factor of $\frac{1}{288}$ followed from opened-closed string duality condition and differs from \eqref{res2}.  Modeling confinement as a Casimir effect appear for example in \cite{FGK}, where $L^{-1}$ proportional correction appeared to a linear potential.

The current model of NC QM lead to exact results (and not only perturbative) in many cases before, see \cite{us1, us2, us3}, mainly because it does not spoil most of symmetries of ordinary QM. Yet, being able to sum over possible frequencies was not guaranteed. 

It shall be interesting to study 3 dimensional Casimir effect in a NC space, possibly with different geometrical shapes (cube, cylinder or a sphere). The presented work firstly focused on the one dimensional case only for its simplicity, but the results displayed similarities to different physical phenomena and should hold in general, possibly in a different form. 

\subsection*{Acknowledgment}

We would like to thank Veronika G\'aliko\'a for her contribution on NC plane waves. This work was partially supported by COST Action MP1405.

\end{document}